\documentclass[aps,pra,twocolumn]{revtex4-2}
\usepackage{color}
\usepackage{longtable}
\usepackage{dcolumn}
\usepackage{graphicx}
\usepackage{bm}
\usepackage{bbm}
\usepackage{times}
\usepackage{nicefrac}
\usepackage{amsmath}
\usepackage{amsfonts}
\usepackage{amssymb}
\usepackage{amsthm}

\begin{document}
\preprint{Version 1.0}

\title{Nuclear recoil correction to the hyperfine splitting in atomic systems}

\author{Krzysztof Pachucki}
\email[]{krp@fuw.edu.pl}
\homepage[]{www.fuw.edu.pl/~krp}

\affiliation{Faculty of Physics, University of Warsaw,
             Pasteura 5, 02-093 Warsaw, Poland}

\date{\today}
\begin{abstract}
We consider leading $O(m/M)$ nuclear recoil corrections to the hyperfine splitting in light atomic systems.
Due to the singularity of hyperfine interactions, the electromagnetic form factors of the nucleus have to be introduced
as a function of the four-momentum square, which goes beyond the Hamiltonian description of the system.
We present a compact formula  for $O(Z\,\alpha)$ recoil corrections and indicate the importance, 
especially for light atoms,  of a more realistic description of the nuclear structure effects.
\end{abstract}


\maketitle

\section{Introduction}

The hyperfine structure (hfs) in atomic systems originates from the coupling between the nuclear and electronic angular momenta.
It is usually measured with many digits' accuracy, for example  $A(^7{\rm Li}) =  401.752 043 3(5)$ MHz \cite{7Li_hfs}. 
Calculations of the hfs are limited in accuracy mostly due to the incompleteness of the basis sets for the multi-electron wave function, 
due to the omission of higher-order QED effects, but also due to the lack of knowledge of the magnetic moment and electric charge distributions within the nucleus.
For example, for $^7$Li  the charge and magnetic moment distribution give the correction to hfs of  $\delta A/A = -369(23)$ ppm \cite{7Li_BW}, 
which dominates theoretical uncertainty \cite{7Li_theory} and is orders of magnitude larger than the experimental one.
Especially the magnetic form factor is believed to be the main source of theoretical uncertainties. 
Various nuclear models, like the single-particle one \cite{single_BW}, have been introduced for a more realistic description 
of the nuclear magnetic interaction. It is still, however, very difficult to estimate uncertainty coming from all these models, 
and strange discrepancies with measurements are observed \cite{7Li_theory}.  

In this work we deal with another, not well studied contribution to the atomic hyperfine structure, which comes from 
the finite nuclear mass, including the second order in the nuclear magnetic moment corrections. It is a small effect,
but nevertheless should be considered in light elements and especially in  muonic atoms. Let us concentrate on hydrogenic systems,
because the extension to many electron atoms is simple. 
The hyperfine splitting for a nonrelativistic hydrogen-like system  in an $S$-state  is given by the Fermi contact interaction 
(in units $\hbar=c=1, e^2=4\,\pi\,\alpha$)
\begin{align}
E_{\rm F} =&\  -\frac{2}{3}\,\langle\psi| \vec\mu\cdot\vec\mu_e\,\delta^3(r)|\psi\rangle \nonumber \\
=&\ \frac{Z\,e^2}{3}\,\frac{\psi^2(0)}{M\,m}\,g\,\vec I\cdot\vec s \nonumber \\
=&\ \frac{8\,\pi}{3}\,Z\,\alpha\,\frac{\psi^2(0)}{M\,m}\,(1+\kappa)\,\vec I\cdot\vec s \,,\label{01}
\end{align}
where $g=2\,(1+\kappa)$, and $\vec\mu$ and $\vec\mu_e$ are the nuclear and electron magnetic moments,  respectively.
We have introduced the (nonstandard) nuclear $g$-factor, similarly to the definition for elementary particles
\begin{align}
\vec \mu =&\ \frac{Z\,e}{2\,M}\,g\,\vec I\,,\label{02}
\end{align}
where $Z\,e$ and $M$ are the charge and mass of the nucleus. For
electrons  the $g$-factor is close to $2$, with a small anomaly
$\kappa_e \approx \alpha/(2\,\pi)$, which is neglected in Eq. (\ref{01}). 
We note that the leading-order $O(Z\,\alpha)$ relativistic correction 
vanishes for  a point-like and infinitely heavy nucleus \cite{eides}. 
For the finite size nucleus the $O(Z\,\alpha)$ correction $E_{\rm Z}$ in the nonrecoil limit is given by \cite{Zemach}
\begin{align}
\frac{E_{\rm Z}}{E_{\rm F}} =&\ \frac{2\,Z\,\alpha\,m_r}{\pi^2}\,
\int\frac{d^3k}{k^4}\,\biggl[\frac{G_E(k^2)\,G_M(k^2)}{1+\kappa}-1\biggr] \nonumber \\
=&\  -2\,Z\,\alpha\, m_r\,r_{\rm Z}
\,,\label{03}
\end{align}
where  $G_E$ and $G_M$ are the electric and magnetic form factors of the nucleus, with normalization $G_M(0) = 1+\kappa$, and $m_r$ is the reduced mass.
It was convenient to rewrite this correction in terms of the Zemach radius $r_{\rm Z}$ which represents the size of the nucleus
\begin{equation}
r_{\rm Z} = \int d^3 r_1 \int d^3 r_2\,\rho_E(r_1)\,\rho_M(r_2)\,|\vec r_1-\vec r_2|, \label{04}
\end{equation}
with $\rho_E$ and $\rho_M$ being the Fourier transforms of $G_E$ and $\tilde G_M = G_M/(1+\kappa)$.
The extension to many electron systems goes as follows. The Fermi contact interaction becomes
\begin{align}
E_{\rm F}  =&\ \frac{8\,\pi}{3}\,\frac{Z\,\alpha}{M\,m}\,(1+\kappa)\,\vec I\cdot \sum_a\langle \psi| \vec s_a\,\delta^{(3)}(r_a)|\psi\rangle \,,\label{xx}
\end{align}
where $\psi$ is a few electron wave function to be calculated numerically. However,  the ratio $E_Z/E_F$ and all other ratios presented in this work
are independent on the electron wave function, therefore  are valid for one- and many-electron atoms and ions.

Below we consider the nuclear recoil correction $\delta E_{\rm Z}$, which is enhanced
${\delta E_{\rm Z}}/{E_{\rm F}} \sim Z\,\alpha\,\ln(m\,r_Z)\,m_r/M$ by the presence of logarithmic terms.
In the first step, however, we investigate a part of the nuclear recoil correction, 
which comes from the second order hyperfine interaction, because the QED theory predictions are different from those obtained from the Dirac equation.
This problem has already been considered in the literature \cite{sternheim, foldy:64, gregson:70, pyykko:72} without obtaining a definite result.

\section{Double hfs contribution}
In the relativistic formalism the leading hyperfine splitting for the point and infinitely heavy nucleus is obtained from the expectation value
\begin{align}
E_{\rm hfs} =  -\langle \psi| e\,\vec\alpha\cdot\vec A |\psi\rangle\,, \label{05}
\end{align}
where
\begin{align}
e\,\vec A(\vec r) = \frac{e}{4\,\pi}\,\vec\mu\times\frac{\vec r}{r^3}\,, \label{06}
\end{align}
and $\psi$ is an eigenstate of the Dirac Hamiltonian
\begin{align}
H = \vec\alpha\cdot \vec p + \beta\,m -\frac{Z\,\alpha}{r}\,. \label{07}
\end{align}
The second order in the magnetic field  correction to the hyperfine splitting is singular; therefore, we have to account for
the finite nuclear size. However, hfs obtained from  the QED theory gives a different result 
from the standard perturbative theory applied to the Dirac equation due to the lack of the crossed diagram. 
This problem has already been considered in several works \cite{foldy:64, gregson:70, pyykko:72}, but none of them calculated this correction correctly. 
Here, we describe this difference in detail and present a result in terms of a specific logarithmic radius. 

Let us consider a QED correction to the binding energy, which is due to the two-photon exchange between the bound electron 
and the nucleus magnetic moment $\vec \mu$, taking into account the nuclear magnetic form factor $\tilde G_M$ with normalization $\tilde G_M(0) = 1$
\begin{widetext}
\begin{align}
\delta E =&\
i\,e^2\,\int\frac{d\,\omega}{2\,\pi}\,\int\frac{d^3k_1}{(2\,\pi)^3}\,\int\frac{d^3k_2}{(2\,\pi)^3}\,
\frac{\tilde G_M(k_1^2-\omega^2)}{\omega^2-k_1^2+i\,\epsilon}\,
\frac{\tilde G_M(k_2^2-\omega^2)}{\omega^2-k_2^2+i\,\epsilon}
\nonumber \\ & \times
\biggl\langle\bar\psi\biggl|\gamma^i\,e^{i\,\vec k_1\vec r}\,
\frac{1}{\not\!p-\gamma^o\,V+\gamma^0\,\omega-m + i\,\epsilon}\,\gamma^j\,e^{-i\,\vec k_2\vec r}\,\biggr|\psi\biggr\rangle
\nonumber \\ & \times
\biggl[ (\vec\mu\times\vec k_1)^{\,i}\,\frac{1}{-\omega+i\,\epsilon}\,(\vec\mu\times\vec k_2)^{\,j}
+ (\vec\mu\times\vec k_2)^{\,j}\,\frac{1}{\omega+i\,\epsilon}\,(\vec\mu\times\vec k_1)^{\,i}
\biggr]\,. \label{08}
\end{align}
The nuclear Hamiltonian is assumed to be $H_{\rm N} = -\vec\mu\cdot\vec B$ with the nuclear kinetic energy neglected. 
Therefore, the above correction is the sum of the ladder and crossed diagrams in the infinite nuclear mass limit.
The Dirac result is obtained by changing the order of the second product of nuclear magnetic moments in the above, namely
\begin{align}
(\vec\mu\times\vec k_2)^{\,j}\,(\vec\mu\times\vec k_1)^{\,i} = 
(\vec\mu\times\vec k_1)^{\,i} \,(\vec\mu\times\vec k_2)^{\,j} +  [(\vec\mu\times\vec k_2)^{\,j}\,,\,(\vec\mu\times\vec k_1)^{\,i}]\,. \label{09}
\end{align}
After dropping the commutator, the two diagrams can be combined
\begin{align}
\delta E_D =&\
i\,e^2\,\int\frac{d\,\omega}{2\,\pi}\,
\biggl(\frac{1}{-\omega+i\,\epsilon} +  \frac{1}{\omega+i\,\epsilon}\biggr)
\int\frac{d^3k_1}{(2\,\pi)^3}\,\int\frac{d^3k_2}{(2\,\pi)^3}\,
\frac{\tilde G_M(k_1^2-\omega^2)}{\omega^2-k_1^2+i\,\epsilon}\,
\frac{\tilde G_M(k_2^2-\omega^2)}{\omega^2-k_2^2+i\,\epsilon}
\nonumber \\ & \times 
\biggl\langle\bar\psi\biggl|\gamma^i\,e^{i\,\vec k_1\vec r}\,
\frac{1}{\not\!p-\gamma^o\,V+\gamma^0\,\omega-m + i\,\epsilon}\,\gamma^j\,e^{-i\,\vec k_2\vec r}\,\biggr|\psi\biggr\rangle
(\vec\mu\times\vec k_1)^{\,i}\,(\vec\mu\times\vec k_2)^{\,j}\,. \label{10}
\end{align}
Taking into account that
\begin{equation}
i\,\int\frac{d\,\omega}{2\,\pi}\,\biggl(\frac{1}{-\omega+i\,\epsilon} +  \frac{1}{\omega+i\,\epsilon}\biggr) f(\omega) = f(0)\,, \label{11}
\end{equation}
we obtain
\begin{align}
\delta E_D =&\
e^2\,\int\frac{d^3k_1}{(2\,\pi)^3}\,\int\frac{d^3k_2}{(2\,\pi)^3}\,
\frac{\tilde G_M(k_1^2)}{k_1^2}\,
\frac{\tilde G_M(k_2^2)}{k_2^2}\,
\biggl\langle\bar\psi\biggl|\gamma^i\,e^{i\,\vec k_1\vec r}\,
\frac{1}{\not\!p-\gamma^o\,V-m}\,\gamma^j\,e^{-i\,\vec k_2\vec r}\,\biggr|\psi\biggr\rangle
(\vec\mu\times\vec k_1)^{\,i}\,(\vec\mu\times\vec k_2)^{\,j} \nonumber  \\ =&\
e^2\,\biggl\langle\bar\psi\biggl|\vec\gamma\cdot\vec A
\frac{1}{\not\!p-\gamma^o\,V-m}\,\vec\gamma\cdot\vec A\,\biggr|\psi\biggr\rangle \label{12}
\end{align}
where
\begin{align}
\vec A(\vec r) =&\ \int\frac{d^3k}{(2\,\pi)^3}\, e^{i\,\vec k\vec r}\,(-i)\,\vec\mu\times\vec k\,\frac{\tilde G_M(k^2)}{k^2}
= \frac{1}{4\,\pi}\,\vec\mu\times\biggl[\frac{\vec r}{r^3}\biggr]_{\rm fs} \label{13}
\end{align}
which in the point nuclear limit  is given by Eq. (\ref{06}).
It is clear that the second order in the nuclear magnetic field correction is not accounted for properly 
by the Dirac equation, and there is an additional contribution from the commutator in Eq. (\ref{09}), 
which was first noticed by Sternheim \cite{sternheim} in 1962.

Let us now return to Eq. (5) and decompose the product of the nuclear magnetic moments into irreducible parts, namely
\begin{align}
\mu^a\,\mu^b = \frac{\delta^{ab}}{3}\,\vec\mu^{\,2} + 
\frac{1}{2}\,\biggl(\mu^a\,\mu^b + \mu^b\,\mu^a-\frac{2\,\delta^{ab}}{3}\,\vec\mu^{\,2}\biggr) +
\frac{1}{2}\,[\mu^a\,,\,\mu^b]\,. \label{14}
\end{align}
The corresponding decomposition of $\delta E$ is
\begin{align}
\delta E = \delta E_S + \delta E_T + \delta E_Z\,. \label{15}
\end{align}
The scalar part $\delta E_S$ shifts the energy, the tensor part $\delta E_T$ shifts the quadrupole hyperfine constant $B_J$, and the vector part $\delta E_Z$
shifts the hyperfine constant $A_J$. The first two terms, the scalar and the tensor, are symmetric with respect to the exchange of indices;
therefore, the formula from the Dirac equation holds for these parts
\begin{align}
\delta E_S =&\ \biggl(\frac{e}{4\,\pi}\biggr)^2\,\biggl\langle\bar\psi\biggl|\vec\gamma\times\biggl[\frac{\vec r}{r^3}\biggr]_{\rm fs}
\frac{1}{\not\!p-\gamma^o\,V-m}\,\vec\gamma\times\biggl[\frac{\vec r}{r^3}\biggr]_{\rm fs}\biggr|\psi\biggr\rangle\,\frac{\vec \mu^{\,2}}{3}\,, \label{16}
\\
\delta E_T =&\ \biggl(\frac{e}{4\,\pi}\biggr)^2\,\biggl\langle\bar\psi\biggl|\biggl(\vec\gamma\times\frac{\vec r}{r^3}\biggr)^a
\frac{1}{\not\!p-\gamma^o\,V-m}\,\biggl(\vec\gamma\times\frac{\vec r}{r^3}\biggr)^b\biggr|\psi\biggr\rangle\,
\frac{1}{2}\,\biggl(\mu^a\,\mu^b - \mu^b\,\mu^a-\frac{\delta^{ab}}{3}\,\vec\mu^{\,2}\biggr)\,, \label{17}
\end{align}
while for the vector part, QED theory gives
\begin{align}
\delta E_Z = &\
i\,e^2\,\frac{[\mu^i\,,\,\mu^j]}{2}\,\int\frac{d\,\omega}{2\,\pi}\,\int\frac{d^3k_1}{(2\,\pi)^3}\,\int\frac{d^3k_2}{(2\,\pi)^3}\,
\frac{\tilde G_M(k_1^2-\omega^2)}{\omega^2-k_1^2+i\,\epsilon}\,
\frac{\tilde G_M(k_2^2-\omega^2)}{\omega^2-k_2^2+i\,\epsilon}\,
\biggl(\frac{1}{-\omega+i\,\epsilon} - \frac{1}{\omega+i\,\epsilon}\biggr)
\nonumber \\ &\ 
\biggl\langle\bar\psi\biggl|(\vec\gamma\times\vec k_1)^i\,e^{i\,\vec k_1\vec r}\,
\frac{1}{\not\!p-\gamma^o\,V+\gamma^0\,\omega-m + i\,\epsilon}\,(\vec\gamma\times\vec k_2)^j\,e^{-i\,\vec k_2\vec r}\,\biggr|\psi\biggr\rangle\,. \label{18}
\end{align}
Let us estimate $\delta E_Z$ by its leading term in the $Z\,\alpha$ expansion, namely
\begin{align}
\delta E_Z \approx &\
i\,e^2\,\frac{[\mu^i\,,\,\mu^j]}{2}\,\psi^2(0)\int\frac{d\,\omega}{2\,\pi}\,\int\frac{d^3k}{(2\,\pi)^3}\,
\biggl[\frac{\tilde G_M(k^2-\omega^2)}{\omega^2-k^2+i\,\epsilon}\biggr]^2\,
\biggl(\frac{1}{-\omega+i\,\epsilon} - \frac{1}{\omega+i\,\epsilon}\biggr)\,M^{ij}\,, \label{19}
\end{align}
where 
\begin{align}
M^{ij}=&\
\biggl\langle\bar u(p)\biggl|(\vec\gamma\times\vec k)^i\,
\frac{1}{\not\!p + \not\!k-m + i\,\epsilon}\,(\vec\gamma\times\vec k)^j\,\biggr| u(p)\biggr\rangle 
=
\frac{\omega}{(m+\omega)^2-k^2-m^2+i\,\epsilon}\,\frac{k^2}{3}\,i\,\epsilon^{ijk}\,\sigma^k \label{20}
\end{align}
\end{widetext}
and $p = (m,0,0,0)$. 
After performing Wick rotation $\omega = i\,w$ with $Q = (w,\vec k)$,  $\delta E_Z$ becomes
\begin{align}
\delta E_Z = &\
i\,e^2\,[\mu^i\,,\,\mu^j]\,\epsilon^{ijk}\,\sigma^k\,\psi^2(0)\,\frac{1}{(2\,\pi)^4}\,\frac{4\,\pi^2}{3}\, C\,, \label{22}
\end{align}
where
\begin{align}
C=&\ \frac{1}{\pi}\,\,\int d\,w\,\int dk\,\frac{\tilde G^2_M(Q^2)}{Q^4}\, \frac{k^4}{2\,m\,i\,w-Q^2}\,. \label{23}
\end{align}
Let $Q$ denote also the modulus of a four-vector $Q$, then after changing the variables $w = \cos(\phi)\,Q$, $k = \sin(\phi)\,Q$ we get
\begin{align}
C = &\ -\frac{1}{\pi}\,\int_0^\infty dQ\,\tilde G^2_M(Q^2) \int_0^\pi d\phi\,\frac{\sin^4(\phi)}{Q-2\,m\,i\,\cos(\phi)}\nonumber \\ =&\
\int_0^\infty \frac{dQ}{Q}\,\tilde G^2_M(Q^2)\,
\biggl(\frac{1}{2\,x^2}-\frac{1}{x}\biggr)\,, \label{24}
\end{align}
where  $x=\sqrt{a^2+1}+1$, and $a=2\,m/Q$.
The integral $C$ is split into two parts, taking into account that the product of
the electron mass $m$ and the finite nuclear size is a small parameter, namely
\begin{align}
C = \int_0^\infty \frac{dQ}{Q}\ldots = \int_0^{\epsilon} \frac{dQ}{Q}\ldots + \int_{\epsilon}^\infty \frac{dQ}{Q}\ldots = C_1+C_2\,. \label{25}
\end{align}
In the first part, we can neglect $q^2$ in the arguments $G_M$, while in the second part we can neglect $m$, so
\begin{align}
C_1 = &\ \int_0^\epsilon  \frac{dQ}{Q}\, \biggl(\frac{1}{2\,x^2}-\frac{1}{x}\biggr) 
= \frac{3}{8} \ln \left(\frac{m}{\epsilon}\right)-\frac{9}{32} \label{26}\\
C_2 = &\ -\frac{3}{8}\,\int_{\epsilon}^\infty \frac{dQ}{Q}\,\tilde G^2_M(Q^2)
= \frac{3}{8}\,\bigl[ \gamma -1 +\ln\epsilon + \ln r_{M^2} \bigr], \label{27}
\end{align}
where
\begin{align}
\ln r_{M^2} = &\ \int d^3 r\int d^3r' \,\rho_M(\vec r)\,\rho_M(\vec r{\,'})\,\ln|\vec r-\vec r{\,'}| \label{28}
\end{align}
and the magnetic moment distribution $\rho_M(\vec r)$ is a Fourier transform of $\tilde G_M(Q^2)$.
 Therefore, we obtain for $C$
\begin{align}
C = \frac{3}{8}\,\Bigl[ \ln (m\,r_{M^2}) + \gamma -\frac{7}{4}  \Bigr] \label{29}
\end{align}
and for $\delta E_Z$
\begin{align}
\delta E_Z = &\
i\,e^2\,[\mu^i\,,\,\mu^j]\,\epsilon^{ijk}\,\sigma^k\,\psi^2(0) \frac{1}{32\,\pi^2}
\Bigl[ \ln (m\,r_{M^2}) + \gamma -\frac{7}{4}  \Bigr]. \label{30}
\end{align}
The nuclear magnetic moment
\begin{align}
\vec\mu = \mu\,\frac{\vec I}{I} = \frac{\mu}{\mu_{\rm N}}\,\frac{\vec I}{I}\,\mu_{\rm N} \label{31}
\end{align}
is usually expressed in terms of the nuclear magneton $\mu_{\rm N}=e/(2\,m_p)$, where $m_p$ is the proton mass, so
\begin{align}
\delta E_Z =& -\Bigl(\frac{\mu}{\mu_{\rm N}\,I}\Bigr)^2 \vec I\cdot\vec\sigma\,\frac{\alpha^2}{4\,m_p^2}\,\psi^2(0)
\Bigl[ \ln (m\,r_{M^2}) + \gamma -\frac{7}{4}  \Bigr] \label{32}
\end{align}
The leading-order hyperfine interaction $E_F$ from Eq. (\ref{01}) is
\begin{align}
E_F = -\frac{2}{3}\,\vec\mu_e\cdot\vec\mu\,\psi^2(0) = \frac{2}{3}\,\frac{\mu}{\mu_{\rm N}\,I}\,\vec I\cdot\vec\sigma\,\frac{\alpha}{m\,m_p}\,\pi\,\psi^2(0); \label{33}
\end{align}
therefore,
\begin{align}
\frac{\delta E_Z}{E_F} =&\ -\frac{\alpha}{\pi}\,\frac{\mu}{\mu_{\rm N}\,I}\,\frac{m}{m_p}\,\frac{3}{8}\,
\Bigl[ \ln (m\,r_{M^2}) + \gamma -\frac{7}{4}  \Bigr] \,.\label{34}
\end{align}
Let us estimate this correction for $^7$Li: $I=3/2$, $\mu/\mu_{\rm N} = 3.256$, $r_{M^2}\approx r_{\rm Z} = 3.25(3)$ fm \cite{7Li_theory}.
The result $\delta E_Z/E_F = 6.1\cdot10^{-6}$ is much larger than the experimental precision $1.2\cdot10^{-9}$, but smaller than the uncertainty
from the not well known Bohr-Weisskopf effect $E_Z/E_F = -369(23)\cdot10^{-6}$. 
However,  it is not the complete recoil correction. This is considered in the following section.

\section{Complete nuclear recoil correction}
In the leading  order in the fine structure constant $\alpha$ the nuclear mass dependence is fully accounted for in $E_F$ in Eq. (\ref{01}).
Here we will consider the nuclear recoil correction in the next order in $\alpha$, and partially follow the previous derivation for the muonic hydrogen \cite{muH}.
The total $(Z\,\alpha)\,E_F$ correction can be represented 
by the two-photon exchange forward scattering amplitude, which in the temporal gauge $A^0=0$
takes the form  \cite{khriplovich, vec_pol}
\begin{widetext}
\begin{equation}
\delta E_{\rm hfs} = \frac{i}{2}\,\int\frac{d\omega}{2\,\pi}\,
\int\frac{d^3k}{(2\,\pi)^3}\,\frac{1}{(\omega^2-k^2)^2}\,
\biggl(\delta^{ik}-\frac{k^i\,k^k}{\omega^2}\biggr)\,
\biggl(\delta^{jl}-\frac{k^j\,k^l}{\omega^2}\biggr)\,
t^{ji}\,T^{kl}\,\psi^2(0)\,,\label{35}
\end{equation}
where 
\begin{eqnarray}
t^{ji} &=& e^2\biggl[
\langle\bar u(p)|\gamma^j\frac{1}{\not\!p\;- \not\!k-m}\,\gamma^i|u(p)\rangle +
\langle\bar u(p)|\gamma^i\frac{1}{\not\!p\;+ \not\!k-m}\,\gamma^j|u(p)\rangle
\biggr] \nonumber \\
&=& i\,\epsilon^{ijk}\,4\,e^2\,\omega
\,s^k\,\frac{(\omega^2-k^2)}
{(\omega^2-2\,m\,\omega-k^2)\,(\omega^2+2\,m\,\omega-k^2)}\,,\label{36}
\end{eqnarray}
and $p$ is the momentum at rest $p=(m,\vec 0)$. The two-photon exchange correction to the hyperfine splitting becomes
\begin{align}
\delta E_{\rm hfs} =&\  -2\,e^2\,\psi^2(0)\,\int\frac{d\,\omega}{2\,\pi}\,
\int \frac{d^3k}{(2\,\pi)^3}\,
\frac{\bigl(\omega^2\,\epsilon^{klj}+k^i\,k^k\,\epsilon^{lij}
-k^i\,k^l\,\epsilon^{kij}\bigr)
\,s^j\,T^{kl}}{\omega\,(\omega^2-k^2)\,(\omega^2-k^2-2\,m\,\omega)\,(\omega^2-k^2+2\,m\,\omega)}\,.\label{37}
\end{align} 
$T^{kl}$ is the corresponding virtual Compton scattering amplitude of the nucleus. 
We will assume at the beginning that the nuclear spin is $S=1/2$, and later show that the $1/M$ correction does not depend on $S$.
The generalization of $t^{kl}$ for an arbitrary $S=1/2$ particle is 
\begin{align}
T^{kl} = (Z\,e)^2\biggl[
\langle\bar u(p)|\Gamma^k(k)\frac{1}{\not\!p\;- \not\!k-m}\,\Gamma^l(-k)|u(p)\rangle +
\langle\bar u(p)|\Gamma^l(-k)\frac{1}{\not\!p\;+ \not\!k-m}\,\Gamma^k(k)|u(p)\rangle\biggr]\,, \label{38}
\end{align}
where
\begin{align}
\Gamma^\mu(k) =&\ \gamma^\mu\,F_1 + \frac{i}{2\,M}\,\sigma^{\mu\nu}\,k_\nu\,F_2 \,. \label{39}
\end{align}
$F_1$ and $F_2$  are related to the electric and magnetic form factors by
\begin{align}
G_E(Q^2) =&\ F_1(Q^2) - \frac{Q^2}{4\,M^2}\,F_2(Q^2)\,, \label{40}\\
G_M(Q^2) =&\ F_1(Q^2)+F_2(Q^2)\,, \label{41}
\end{align}
with $Q^2 = k^2-\omega^2$. We  assume that the nucleus is described exclusively by electromagnetic form factors and thus neglect inelastic contributions.
In addition, we neglect also higher order in $1/M$ terms and obtain
\begin{align}
T^{ji} =&\ i\,\epsilon^{ijk}\,4\,(Z\,e)^2\,\omega
\,\frac{\Bigl[\omega^2\,I^k\,(2\,G_M-G_E)\,G_E -k^2\,I^k\,G_M\,G_E - k^k\,(\vec k\cdot\vec I)\,G_M\,(G_M-G_E)\Bigr]}
{(-2\,M\,\omega-k^2)\,(2\,M\,\omega-k^2)}. \label{42}
\end{align}
Although $T^{ji}$ is derived for $I=1/2$, the formula is valid for an arbitrary spin, and the proof is as follows.
Let us consider a Hamiltonian for an arbitrary spin $I$ particle, including up to $1/M^2$ terms, namely
\begin{equation}
H = \frac{\vec\Pi^2}{2\,M} + Z\,e\,A^0 -\frac{Z\,e}{2\,M}\,g\,\vec I\cdot\vec B
-\frac{Z\,e}{4\,M^2}\,(g-1)\,\vec I\cdot[\vec E\times\vec\Pi-\vec\Pi\times\vec E],\label{43}
\end{equation}
where $\vec\Pi = \vec P-Z\,e\,\vec A$.
Electromagnetic  form factors are neglected in the above, because they cannot be included on the Hamiltonian level.
From this Hamiltonian, following  Refs. \cite{khriplovich, vec_pol}, we derive  the following scattering amplitude
\begin{align}
T^{ji} =&\ i\,\epsilon^{ijk}\,4\,(Z\,e)^2\,\omega
\,\frac{\bigl[\omega^2\,I^k\,(g-1) -k^2\,I^k\,g/2 - k^k\,(\vec k\cdot\vec I)\,g\,(g-2)/4\bigr]}
{(-2\,M\,\omega-k^2)\,(2\,M\,\omega-k^2)}\,,\label{44}
\end{align}
which coincides with Eq. (\ref{42})  with $G_E(0)=1$ and $G_M(0)=g/2$.
Therefore, Eq. (\ref{42}) is valid for an arbitrary spin finite size particle. 

The resulting correction to {\em hfs} can now be written as
\begin{align}
\delta E_{\rm hfs} =&\ -\frac{16\,i}{3}\,(Z\,e^2)^2\,\psi^2(0)\,\vec I\,\vec s\,\int\frac{d\omega}{2\,\pi}\,
\int\frac{d^3k}{(2\,\pi)^3}\nonumber \\ &\times
\frac{ G_E^2\,\omega^2(2\,k^2 - 3\,\omega^2) + 2\,G_E\,G_M\,(k^4 - 3\,k^2\,\omega^2 + 3\,\omega^4)-G_M^2\,k^2\,\omega^2 }{ (\omega^2-k^2)
(\omega^2-2\,m\,\omega-k^2)\,(\omega^2+2\,m\,\omega-k^2)(-2\,M\,\omega-k^2)\,(2\,M\,\omega-k^2)  }\,. \label{45}
\end{align}
Because form factors are functions of $\omega^2-k^2$, we perform at first 
the angular integration in the four-dimensional space as in the previous section, and then perform $1/M$ expansion
\begin{eqnarray}
\delta E_{\rm hfs} & = & -\frac{16}{3}\,(Z\,\alpha)^2\,\psi^2(0)\, 
\vec I\,\vec s\,\int\frac{dQ}{Q}\,\biggl(\frac{T^{(1)}}{M+m} + \frac{T^{(2)}}{M^2} +\ldots\bigg) = E_Z + \delta E_Z\,, \label{46}
\end{eqnarray}
where  
\begin{equation}
T^{(1)} = \frac{4}{Q}\,\left(G_M(0)\,G_E(0)-G_M(Q^2)\,G_E(Q^2)\right) \label{47}
\end{equation}
coincides with $E_Z$ in Eq. (\ref{03}), and
\begin{align}
T^{(2)} = &\  
 \left(\frac{2}{x'}-\frac{2}{x}-\frac{1}{x^2}\right)\,G_E\left(Q^2\right)\,G_M\left(Q^2\right)   
    +\left(\frac{2}{x}+\frac{1}{2\,x^2}\right)
    G^2_E\left(Q^2\right)+\left(\frac{1}{2\,x^2}-\frac{1}{x}\right)
    G^2_M\left(Q^2\right)\,,
\end{align}
where $x = \sqrt{a^2+1}+1$,   $x' = \sqrt{a^2+1}+a$, and $a=2\,m/Q$. 
The term with $G_M^2$ is in agreement with the result for the second-order magnetic interaction Eq. (\ref{24}).

In order to perform the last integral, we explore the smallness of the electron mass and split it into two parts
\begin{align}
C = \int_0^\infty \frac{d Q}{Q}\,T^{(2)} = \int_0^{\epsilon} \frac{d Q}{Q}\,T^{(2)} + \int_{\epsilon}^\infty \frac{d Q}{Q}\,T^{(2)} = C_1+C_2\,,
\end{align}
with $\epsilon\rightarrow 0$ after expansion in $m$.
In the first part, we can neglect $q^2$ in the arguments of $G_E$ and $G_M$
\begin{align}
T^{(2)} \approx&\ 
\frac{2}{x}+\frac{1}{2\,x^2}
+\left(\frac{2}{x'}-\frac{2}{x}-\frac{1}{x^2}\right)\,G_M(0)
+ \left(\frac{1}{2\,x^2}-\frac{1}{x}\right)\, G_M(0)^2\,,
\end{align}
and the integral is
\begin{align}
C_1 =&\ \int_0^{\epsilon} \frac{d Q}{Q}\,T^{(2)} = \Bigl[\frac{15}{32} - \frac{9}{8}\,\ln\Bigl(\frac{m}{\epsilon}\Bigr)\Bigr]
+ G_M(0)\,\Bigl[-\frac{39}{16} - \frac{3}{4}\,\ln\Bigl(\frac{m}{\epsilon}\Bigr)\Bigr]  + G_M(0)^2\,\Bigl[-\frac{9}{32} + \frac{3}{8}\,\ln\Bigl(\frac{m}{\epsilon}\Bigr)\Bigr]\,. \label{50}
\end{align}
In the second part we can neglect $m$, so
\begin{align}
T^{(2)} \approx &\ \frac{3}{8}\,\big[3\,G_E(Q^2) - G_M(Q^2)\big]\,\big[G_E(Q^2) + G_M(Q^2)\big]
\end{align}
and
\begin{align}
C_2 =&\ \int_{\epsilon}^\infty \frac{d Q}{Q}\,T^{(2)}  =
-\frac{9}{8}\, \big[\gamma-1+\ln\epsilon+\ln r_{E^2}\big] 
-\frac{3}{4}\,G_M(0)\, \big[\gamma-1+\ln\epsilon+\ln r_{EM}\big]
+\frac{3}{8}\,G^2_M(0)\, \big[\gamma-1+\ln\epsilon+\ln r_{M^2}\big], \label{52}
\end{align}
where $r_{E^2}$ and $r_{EM}$ are defined in analogy to Eq. (\ref{28}).
The total integral $C=C_1+C_2$ is
\begin{align}
C =&
-\frac{9}{8}\, \big[\gamma-\frac{17}{12}+\ln(m\, r_{E^2})\big] 
-\frac{3}{4}\,G_M(0)\, \big[\gamma+\frac{9}{4}+\ln(m\, r_{EM})\big]
+\frac{3}{8}\,G^2_M(0)\, \big[\gamma-\frac{7}{4}+\ln(m\, r_{M^2})\big]\,,
\end{align}
and the corresponding correction to the hyperfine splitting is
\begin{align}
\delta E_Z =&\  -\frac{16}{3}\,\frac{(Z\,\alpha)^2}{M^2}\,\psi^2(0)\, \vec I\,\vec s\;C\,.
\end{align}
The relative correction 
\begin{align}
\frac{\delta E_Z}{E_F} =&\
-\frac{Z\,\alpha}{\pi}\,\frac{m}{M}\,\frac{3}{8}\,\biggl\{
g\,\biggl[\gamma-\frac{7}{4} + \ln(m\, r_{M^2})\biggr]
- 4\,\biggl[\gamma+\frac{9}{4} + \ln(m\,r_{EM})\biggr]  
-\frac{12}{g}\,\biggl[ \gamma-\frac{17}{12} + \ln(m\,r_{E^2})\biggr]
\biggr\}\,, \label{57}
\end{align}
\end{widetext}
is the main result of this work. The nuclear recoil correction is  represented in terms of the $g$-factor defined in Eq. (\ref{02}), 
and effective logarithmic radii $r_{E^2}, r_{M^2}, r_{EM}$,
which are expected to be of the same order as the Zemach radius $r_Z$ in Eq. (\ref{05}). Moreover, this result is independent on the number of electrons,
as long as the expansion in $Z\alpha$ makes sense.
This allows for simple estimation of nuclear finite size recoil effects for an arbitrary element.
For example, for $^7$Li with $g(^7{\rm Li}) = 5.041$  they amount to $\delta E_Z/E_F = 1.8\cdot 10^{-6}$,
which is smaller than the second-order magnetic moment part from the previous section, so cancellation between all three parts is observed.

However, the use of elastic form factors has its limitations. The true  nuclear structure effects might be significantly different,
as is the case of hyperfine splitting in muonic deuterium, where the Zemach correction Eq. (\ref{04}) is of the opposite sign
to the total nuclear structure contribution \cite{muD}. In the next section, we present a more accurate treatment
that goes into more detail regarding the nuclear structure.

\section{Nuclear structure correction to hyperfine splitting}
The use of elastic form factors is very convenient because we can solve the Dirac equation in the modified Coulomb and magnetic potentials, and
for heavy nuclei there is little choice.
The first problem of this approach is the lack of knowledge on the magnetic form factor, i.e., the distribution of the magnetic moment within the nucleus.
Perhaps, for some nuclei, the single-particle model \cite{single_BW}  gives a good estimation of the magnetic form factor, but this is not all;
we also have to estimate the so-called inelastic effects. 
Because there is no reliable nuclear structure theory that includes electromagnetic interactions, we are left with some 
more or less convincing arguments. Let us therefore consider light atoms, for which we can use perturbation theory to account for the nuclear structure effect.
In the zeroth order, the nucleus is a point particle with some mass. The leading nuclear structure effects come from the two-photon exchange diagrams,
see Eq. (\ref{37}). The exchanged photon momentum can be of the order of nuclear excitation energies. Then, such a contribution is associated with the nuclear 
vector polarizability. However, the exchanged photon momentum can be of the order of the inverse nuclear size. For such a photon momentum,
nucleons are seen as individual particles, each one with its own kinetic energy and mutual interactions.
In this case we perform an expansion in the small parameter, which is the ratio of the excitation energy and this photon momentum, 
which leads to the so-called Low formula. Below, following Ref. \cite{vec_pol}, we rederive this formula and show that the leading nuclear recoil effects vanish.
 
Let us split $\delta E_{\rm hfs}$ in Eq. (\ref{39}) into three parts:
\begin{equation}
\delta E_{\rm hfs} = E_{\rm Low} + E_{1\rm nuc} + E_{\rm pol}.
\label{58}
\end{equation}
The first two parts come from the exchanged photon momentum of the order of the inverse of the nuclear size, which is much larger than
the binding energy per nucleon, and the last part is from the photon momentum as small as the nuclear excitation energy.
$E_{\rm Low}$ is the contribution to hfs of order $Z\,\alpha\,m\,r_N$, where $r_N$ is the size of the nucleus, which originates when 
the exchanged photons hit two different nucleons. This contribution was first derived by Low in \cite{low} and has been 
reanalyzed and calculated for such nuclei as D, T, and $^3$He by Friar and Payne in Ref. \cite{friar}. 
The second term in Eq. (\ref{63}) $\delta E_{\rm 1nuc}$ originates when both photons hit the same nucleon. 
It is a sum of the Zemach corrections from all individual nucleons,
\begin{equation}
E_{1\rm nuc} = -\frac{8\,\pi}{3}\,\alpha^2\,\frac{\psi^2(0)}{m_p+m}\,\vec s\cdot \bigl\langle\sum_a g_a\,\vec s_a\,r_{a\rm Z}\bigr\rangle,
\end{equation}
where $r_{pZ} = 0.883(19)$ fm for the proton and $r_{nZ} = 0.102(39)$ fm for the neutron in muonic atoms,
and  $r_{pZ} = 0.863(20)$ and $r_{nZ} = 0.347(38)$ in electronic atoms \cite{tomalak1, tomalak2}.
These effective charge radii include elastic, inelastic, and recoil effects, and they depend on the lepton mass. 
The third correction, due to the nuclear polarizability $E_{\rm pol}$, comes from the exchanged photon momentum of the order
of the nuclear excitation energy. It has been analyzed for electronic atoms in Ref. \cite{vec_pol} and  for muonic ones in Ref. \cite{muD}.
These derivations are quite complicated, and have not yet been confirmed by independent analysis and thus will not be repeated  here.

We now consider the case in which the photon is emitted and absorbed by different nucleons,
denoting the resulting correction as $E_{\rm Low}$.
The Compton amplitude $T^{\mu\nu}$ obeys $k_\mu\,T^{\mu\nu}=0$.
From this, we obtain $k^i\,T^{ik} = \omega T^{0k}$, and use this
to rewrite Eq. (\ref{39})  in the form
\begin{widetext}
\begin{eqnarray}
\delta E_{\rm hfs} &=& -2\,e^2\,\psi^2(0)\,\int\frac{d\,\omega}{2\,\pi}\,
\int \frac{d^3k}{(2\,\pi)^3}\,
\frac{s^j\,\epsilon^{klj}\,\bigl[\omega\,T^{kl}+k^l\,(T^{0k}-T^{k0})\bigr]}
{(\omega^2-k^2)\,(\omega^2-k^2+2\,m\,\omega)\,(\omega^2-k^2-2\,m\,\omega)}.
\label{60}
\end{eqnarray} 
Because the dominating contribution in the above integral comes from
$\omega, k$, of order of the inverse of the nuclear size, which is much smaller
than the nucleon mass, we may use the nonrelativistic
approximation for $T^{\mu\nu}$,
\begin{align}
T^{\mu\nu} =\sum_{a\neq b}\int d^3 r\,d^3 r'\biggl\langle
J^\mu_a(r)\,e^{i\,\vec k\,\vec r}\frac{1}{E-H_{\rm N}-\omega+i\,\epsilon}
J^\nu_b(r')\,e^{-i\,\vec k\,\vec r'} 
+ J^\nu_b(r')\,e^{-i\,\vec k\,\vec r'}\frac{1}{E-H_{\rm N}+\omega+i\,\epsilon}
J^\mu_a(r)\,e^{i\,\vec k\,\vec r}\biggr\rangle.
\label{61}
\end{align}
The contribution coming from $T^{kl}$ in Eq. (\ref{60}),
is of nominal order $Z\,\alpha\,m/m_p\,E_F$; therefore, 
after carrying out the $\omega$ integration we can neglect
$H_{\rm N}-E$ in comparison to $k$. It is then
proportional to the commutator $[J^k_a(r),J^l_b(r)]$,
which vanishes for different nucleons $a\neq b$. The second term
in Eq. (\ref{60}) involving $T^{0j}$ is of nominal order
$Z\,\alpha\,m\,r_N\,E_F$; therefore, wel keep also the second term in the expansion:
\begin{equation}
\frac{1}{k+H_{\rm N}-E} \approx \frac{1}{k} -\frac{H_{\rm N}-E}{k^2}.
\label{62}
\end{equation}
This term includes the kinetic energy of the nucleus, and thus it would give the nuclear recoil correction.
However, this term, after commuting with current operators, leads to the correction proportional to
\begin{align}
\bigl\langle J^0_a(r)\,e^{i\,\vec k\,\vec r}\,(H_{\rm N}-E)\,J^i_b(r')\,e^{-i\,\vec k\,\vec r'}+
J^i_b(r')\,e^{-i\,\vec k\,\vec r'}\,(H_{\rm N}-E)\,J^0_a(r)\,e^{i\,\vec k\,\vec r}\bigr\rangle
= \bigl\langle [J^0_a(r)\,e^{i\,\vec k\,\vec r},[H_{\rm N}-E, J^i_b(r')\,e^{-i\,\vec k\,\vec r'}]]\bigr\rangle,
\label{63}
\end{align}
which also vanishes or is small, because $a\neq b$. Therefore, the recoil corrections vanish,
and $E_{\rm Low}$ after $\omega$ integration becomes
\begin{equation}
E_{\rm Low} = e^2\,\psi^2(0)\,\int_\Lambda \frac{d^3 k}{(2\,\pi)^3}\,
\frac{2\,i}{k^6}\,\vec \sigma\times\vec k\,\sum_{a\neq b}\int d^3 r\,d^3 r'
\bigl\langle J_a^0(r)\,e^{i\,\vec k\,(\vec r-\vec r')}\,\vec J_b(r')\bigr\rangle.
\label{64}
\end{equation}
This result does not depend on the fermion mass, and so it is valid for electronic and muonic atoms, but it is divergent at small $k$.
This linear divergence is eliminated by subtraction, and the low $k$ contribution is included separately as a nuclear polarizability correction \cite{vec_pol}. 
Assuming the nonrelativistic approximation for the interaction of the nucleon with the electromagnetic field,
\begin{equation}
H = \frac{(\vec p-e\,\vec A)^2}{2\,m} + e\,A^0 - \vec\mu\cdot\vec B, \label{66}
\end{equation}
we obtain the ``Low'' nuclear structure correction to the hyperfine splitting \cite{low}
\begin{eqnarray}
E_{\rm Low} &=& e^2\,\psi^2(0)\,\int\frac{d^3k}{(2\,\pi)^3}\,
\frac{2\,i}{k^6}\,\vec\sigma\times\vec k\,\sum_{a\neq b} \frac{e_a\,e_b}{m_b}\,
\biggl\langle e^{i\,\vec k\,\vec r_{ab}}\,
\biggl(\vec p_b-\frac{i\,g_b}{4}\,\vec\sigma_b\times\vec k\biggr)\biggr\rangle
\nonumber \\ &=&
\frac{\alpha}{16}\,\psi^2(0)\,\vec\sigma\,\sum_{a\neq b}
\frac{e_a\,e_b}{m_b}\,\biggl\langle
4\,r_{ab}\,\vec r_{ab}\times\vec p_b +\frac{g_b}{r_{ab}}
\bigl[\vec r_{ab}\,(\vec r_{ab}\cdot\vec\sigma_b)-3\,\vec\sigma_b\,r_{ab}^2 \bigr]\biggr\rangle\,,
\label{67}
\end{eqnarray}
\end{widetext}
in which it is assumed that $e_a\,g_a = e g_n$ for the neutron.
Relativistic effects give corrections of the order of the nucleon binding energy over the nucleon mass, and therefore will be neglected.
Further corrections such as the finite nucleon size, and pion exchange currents have already been presented in Ref. \cite{friar}.

Let us consider the special case of  a spherically symmetric nucleus and neglect the proton-neutron mass difference.
\begin{align}
E_{\rm Low} =& -\frac{8\,\pi}{3}\,\alpha^2\,\frac{\psi^2(0)}{m_n}\,\sum_{a-{\rm protons}}\sum_b\bigl\langle r_{ab}\,g_b\,\vec s_b\bigr\rangle\,\vec s
\label{68}
\end{align}
This formula  looks similar to $E_Z$ Eq. (\ref{03}) with the essential difference being the meaning of nuclear radii: $r_Z$ versus averaged 
internucleon distance $r_{ab}$.

In summary, $E_{\rm Low} + E_{\rm 1nuc}$ may be considered as a more realistic representation of the leading nuclear structure contribution to hfs,  
in comparison to the elastic contribution $E_Z$ in Eq. (\ref{03}). The nuclear recoil corrections corresponding to $E_{\rm Low}$ vanish,
while those to $E_{\rm 1nuc}$ are already included in corresponding 
Zemach radii for the proton and neutron \cite{tomalak1, tomalak2}. 
It is  because they have been calculated from the measured scattering amplitude through dispersion relations.
 
 \section{Conclusions}
The accurate formulation of nuclear structure effects in hyperfine splitting has not yet been performed in the literature. 
If the elastic form factors are used as a starting point,
then the nuclear recoil correction, which is  considered in this work, contributes to hfs at the relative order of $ O(Z\,\alpha\,\ln(r\,m_e)/M)$.
In the particular case of $^7$Li it amounts to about $1.8\times10^{-6}$. 
For all other elements the precise value of the recoil correction $\delta E_Z/E_F$ depends on the 
nuclear magnetic moment and on effective nuclear electromagnetic radii. 
Nevertheless, we expect it to be about $\pm 10^{-6}$ for all other nuclei,
which is usually not significant in comparison to uncertainties due to the little-known magnetic form factor. 
What is interesting, nevertheless, is that for the recoil corrections one has to include electromagnetic form factors of the nucleus as functions of 
not only the exchanged momentum but also the exchanged energy, which goes beyond the Hamiltonian description of the system.

There are further corrections, namely those due to the nuclear polarizability, which are usually overlooked in the literature. 
They are probably not very significant for most of atomic systems, but might be responsible for small discrepancies with measured values, 
like those for $^6$Li \cite{7Li_theory}.    
For light nuclei, we can employ a perturbative approach and consider the nuclear structure effects in powers of the fine structure constant $\alpha$.
The two-photon exchange correction $O(Z\,\alpha)\,E_F$ can be represented, depending on the value of exchanged momenta, 
in terms of three contributions, as in Eq. (\ref{58}).  Their evaluation, however, have not yet been performed, except partially for the deuteron.
It should also be mentioned that three-photon exchange correction $O(Z\,\alpha)^2$ has already been considered \cite{muD}, but without detailed analysis.

We point out, that calculations of the nuclear structure  effects can be verified only if atomic calculations are accurate enough, 
which can be achieved for few-electron systems.
This is the case for hfs in $^6$Li and $^7$Li, in which nuclear structure effects are found to be not consistent 
with those derived from elastic form factors \cite{7Li_theory, 7Li_chinese}. 
The calculated here finite nuclear mass (recoil) corrections are too small to explain these discrepancies. 
Probably the best system with which to study nuclear structure effects
is the hyperfine splitting in muonic deuterium, where $E_Z$ is of the right order of magnitude, but of the opposite sign to the nuclear structure contribution $\delta E_{\rm hfs}$  extracted from the measured hfs \cite{muD}. This is a clear indication, that description of the nucleus through elastic form factors does not always work for the estimation of nuclear effects in the hyperfine splitting.

\acknowledgments
This work was supported by the National Science Center (Poland) Grant No. 2017/27/B/ST2/02459.

\end{document}